# The search for candidate relevant subsets of variables in complex systems

M. Villani[1,2], A. Roli[3], A. Filisetti[1], M. Fiorucci[1,4], Irene Poli[1,4] and R. Serra[1,2,*]

[1] European Centre for Living Technology, Ca' Minich, S. Marco 2940, 30124 Venezia, Italy
[2] Dept. of Physics, Informatics and Mathematics, University of Modena e Reggio Emilia, v. Campi 213b, 41125 Modena, Italy
[3] DISI Alma Mater Studiorum University of Bologna Campus of Cesena, Via Venezia 52, I-47521 Cesena, Italy
[4] Ca' Foscari University of Venice, Department of Environmental Sciences (DAIS)
[*] corresponding author, rserra@unimore.it

**Abstract**

In this paper we describe a method to identify "relevant subsets" of variables, useful to understand the organization of a dynamical system. The variables belonging to a relevant subset should have a strong integration with the other variables of the same relevant subset, and a much weaker interaction with the other system variables. On this basis, extending previous works on neural networks, an information-theoretic measure is introduced, i.e. the Dynamical Cluster Index, in order to identify good candidate relevant subsets. The method does not require any previous knowledge of the relationships among the system variables, but relies on observations of their values in time. We show its usefulness in several application domains, including: (i) random boolean networks, where the whole network is made of different subnetworks with different topological relationships (independent or interacting subnetworks); (ii) leader-follower dynamics, subject to noise and fluctuations; (iii) catalytic reaction networks in a flow reactor; (iv) the MAPK signaling pathway in eukaryotes. The validity of the method has been tested in cases where the data are generated by a known dynamical model and the Dynamical Cluster Index method is applied in order to uncover significant aspects of its organization; however it is important to stress that it can also be applied to time series coming from field data without any reference to a model. Given that it is based on relative frequencies of sets of values, the method could be applied also to cases where the data are not ordered in time. Several indications to improve the scope and effectiveness of the Dynamical Cluster Index to analyze the organization of complex systems are finally given.

## 1. Introduction

In artificial as well as in natural life one often encounters systems that show some forms of organization, that cannot however be understood by referring to simple hierarchical models (like e.g. a tree). In most interesting cases one is faced with complicated situations, sometimes referred to as "tangled hierarchies", where a clear-cut hierarchy of levels, with a unique well-defined direction of information flow cannot be found.

Moreover, in several cases, the interactions among the interesting variables are largely, or at least partly unknown, therefore it is necessary to infer some hypotheses about the organization by observing its behavior "from outside", i.e. like a black box or a grey box.

In this paper we will address the issue of identifying sets of variables that are good candidates as "relevant subsets" for describing the organization of a dynamical system. We will assume that some variables can be observed and that they change in time (possibly reaching an attractor state) and we will look for groups of variables that can represent the required relevant subsets for describing the system organization.

Note that the very notion of a relevant subset is somewhat ill-defined. However this feature is shared by several other interesting concepts, like e.g. that of a cluster. In general, we stress that a good candidate relevant subset (CRS for short) should provide important indications about some key features of the system organization. While this might seem abstract at this stage, we will clarify how it works in a number of examples and then we will come back to the issue of the definition.

The main features of good CRSs can be tentatively identified as follows:
1. The variables that belong to a relevant subset have a strong interaction and integration with the other variables of the same relevant subset
2. They have a weaker integration with the other system variables or relevant subsets

As far as point 2 is concerned, let us stress that a particular case is that of candidate subsets that are essentially uncoupled from the rest of the system. Our methods allow one to identify these sets but the analysis will be mostly focused on more interesting cases where the interaction does not vanish. Note also that CRSs can overlap, i.e. they don't need to be disjoint sets.

The outcome of the analysis we propose here is essentially a list of subsets, ranked according to the above criteria, that provide clues to understand the system organization. The list cannot be used with brute force methods: its application to a particular case requires some care, but it can lead one to discover very interesting non obvious relationships.

A peculiar feature of the method that will be described and applied below is that it does not require knowledge of the system structure and rules. To clarify this statement, consider e.g. the case of a network composed of $N$ nodes; to each node $i$ a dynamical variable $x_i$ is associated, that changes in discrete times ($t$=1,2,…) according to a precise first-order differential (or difference) equation. As usual, we will say that there is a directed link from node $i$ to $k$ if $x_i$ appears on the right hand side of the dynamical equation of $x_k$. The set of dynamical equations and the set of links will be collectively referred to as the dynamical rules and the topology of the network, respectively. It is important to stress that the application of the method only requires knowledge of the behavior in time of the variables $x_1…x_N$, without any prior knowledge of the topology or of the dynamical rules. All that is needed is a set of numerical values of the relevant variables in time, therefore the method can be applied to models as well as to real-world data. If further information concerning the system (e.g., topology) are available, they can be profitably used, as it will be discussed in the final section, but they are not necessary.

An interesting feature of our approach is that it relies upon information theory. While the idea of applying concepts and measures of information theory to the study of complex systems is certainly not new, we found that a measure of this kind, the Cluster Index that was introduced by Tononi and Edelman [Tononi et al. 1998] in the study of neural networks close to a stationary state, can be profitably generalized to study dynamical systems. This generalization is called here the Dynamical Cluster Index (DCI). This measure will be described in detail in section 2, but we want to remark here that it can be applied both to nonlinear systems with different attractors, and to transients; the choice will be dictated by the available information and, when both are available, both can be used.

The DCI can be applied to any subset of the system variables; its evaluation allows us to identify those subsets that are strongly integrated within themselves, and loosely interacting with the rest of the system. While it cannot be claimed that these subsets always correspond to "important" intermediate levels, they are good candidates for that role, so the application of the DCI is a good method to draw attention on hierarchical relationships, when there are some.

Note that the dynamical cluster index has actually been introduced in [Villani et al. 2013]; the present paper describes further developments of the method (see sections 2 and 7) and it also shows new interesting applications (see sections 4, 5 and 6) that highlight its effectiveness and help uncovering some of its features. In order to make the paper self-contained, the DCI is also fully reviewed (section 2) and an application to a model of random boolean networks, that represents a basic testbed for the method itself, is given in section 3.

A methodological aspect concerns the way in which this approach can be tested. As it has already been stressed, it can be applied to data as well as to models, but in order to ascertain its properties it is useful to focus on the latter. We will therefore consider in this work dynamical models, whose role will be twofold: on the one hand, they will generate data concerning the behavior in time of the system variables, that will be used to feed the method that will come out with a list of candidate relevant subsets. On the other hand, the structure of the model is perfectly well-known to us, so we will compare the proposed "organizational hints" with the true organization of the model. In this way we will get information about the reliability of the method, as well as clues concerning the way in which its outcomes should be used to make proper inferences.

In the following, after presenting and discussing the method in section 2, we will show the results of its application to various models. In section 3 the random boolean network framework will be considered, in order to verify that the method is able to identify subsets that make sense. In section 4 a peculiar (artificial) model made of leaders and followers will be introduced, and it will be used to test the response of the system to external changing stimuli and to increasing noise levels.

While the systems considered in sections 3 and 4 are boolean, in the following sections the variables to be analyzed are continuous. Yet, in order to apply the proposed method, a peculiar discrete coding will be used, that does not describe the instantaneous values of the state variables, but rather the signs of their first-order time derivatives. This three-level coding (increasing, constant, decreasing) has proven very effective in analyzing a simulated chemical system in a continuously stirred tank reactor, individuating the relevant chemical subnetworks (described in section 5) as well as in interpreting the results of a model, based on experimental data, of the dynamics of protein phosphorilation-dephosphorilation in the MAPK system, whose details are given in section 6. This is so far the more complex case that has been analyzed with the DCI method.

Finally, in section 7 we will revisit the method, taking into account the lessons learnt from the case studies. In particular, we will show that the DCI is a useful technique, but its application requires some ingenuity, and it should not be considered a brute force method. Some of the problems, to be discussed also in section 7, are related to the fact that good clusters of $k$ variables usually include good clusters of $k$-1 variables, so identifying the "more relevant" subsets may be an ill-defined and controversial problem.

The DCI is only one (actually, the simplest) measure out of a number of different alternatives based on information theory that can be applied to identifying CRSs. Indications for further studies will also be given in the final section.

## 2. The dynamical cluster index

Let us consider a system $U$, composed of $N$ elements assuming finite and discrete values. The cluster index, as defined by Edelman and Tononi [Tononi et al. 1998], is an information theoretical measure based on the Shannon entropy of both the single elements and sets of elements in $U$.
According to information theory, the entropy of an element $x_i$ is defined as:

$$H(x_i) = -\sum_{v \in V_i} p(v) \log(p(v)) \qquad (1)$$

where $V_i$ is the set of the possible values of $x_i$ and $p(v)$ the probability of occurrence of symbol $v$. The entropy of a pair of elements $x_i$ and $x_j$ is defined by means of their joint probabilities:

$$H(x_i, x_j) = -\sum_{v \in V_i} \sum_{w \in V_j} p(v,w) \log(p(v,w)) \qquad (2)$$

Eq. 2 can be extended to sets of $k$ elements considering the probability of occurrence of vectors of $k$ values.

In this work, we deal with observational data, therefore probabilities will be estimated by means of relative frequencies.

The cluster index $C(S)$ of a set $S$ of $k$ elements is defined as the ratio between two other measures, i.e. the *integration I(S)* of $S$ and the *mutual information* between $S$ and the rest of the system $U$-$S$. Let us define the integration[1] as:

$$I(S) = \sum_{x \in S} H(x) - H(S) \qquad (3)$$

I(S) represents the deviation from statistical independence of the $k$ elements in $S$. Its maximum value is $(k-1)\log(b)$, where $b$ is the number of levels of the variable; so the ratio $I/(k-1)$ provides useful information to compare the integration of subsets of different size. This ratio can in some cases directly provide useful indications to identify candidate relevant subsets, however it has been observed that the dynamical cluster index is effective even in cases where the normalized integration is not sufficient. Then, let us define (as usual) the mutual information $M(S;U-S)$ as

$$M(S; U-S) \equiv H(S) + H(S | U-S) = H(S) + H(U-S) - H(S, U-S) \qquad (4)$$

---

[1] The integration is also known as intrinsic information or multi-information

where *H(A|B)* is the conditional entropy and *H(A,B)* the joint entropy. Finally, the cluster index *C(S)* is defined as:

$$C(S) = \frac{I(S)}{M(S;U-S)} \qquad (5)$$

Since C is defined as a ratio, it is undefined in all those cases where *M(S;U-S)* vanishes. In this case, however, the subset S is statistically independent from the rest of the system and it therefore has to be analyzed separately. These cases can be screened in advance. In particular, if *I* is finite while *M*=0, then *C*→+∞ and *S* is to some extent integrated, but also segregated from the rest of the system. It can be an interesting subset, but not of the kind we are looking for.

Note that *C(S)* scales with the size of *S*, therefore the cluster index values of subsystems of different size cannot be in principle compared. To overcome this limitation, cluster index values of systems of different size need to be normalized. To this aim, let us define a reference system, i.e., the homogeneous system $U_h$, randomly generated in accordance with the probability of each single state measured in the original system *U*, along all its series of states. Then, in order to have a reference value for the cluster index for each size, for each subsystem size of $U_h$ the average integration $\langle I_h \rangle$ and the average mutual information $\langle M_h \rangle$ are computed.
Hence, the cluster index value of any subsystem *S* can be normalized by means of the appropriate normalization constant based on the size *S*:

$$C'(S) = \frac{I(S)}{\langle I_h \rangle} \bigg/ \frac{M(S;U-S)}{\langle M_h \rangle} \qquad (6)$$

Furthermore, to assess the significance of the differences observed in the cluster index values, a statistical index $T_c$ is computed:

$$T_c(S) = \frac{C'(S) - \langle C'_h \rangle}{\sigma(C'_h)} = \frac{\nu C - \nu \langle C_h \rangle}{\nu \sigma(C_h)} = \frac{C - \langle C_h \rangle}{\sigma(C_h)} \qquad (7)$$

where $\langle C'_h \rangle$ and $\sigma(C'_h)$ are the average and the standard deviation of the population of normalized cluster indices with the same size of *S* from the homogeneous system and $\nu=\langle M_h\rangle/\langle I_h\rangle$ is the normalization constant. Note that in Eq. 7 we prove that, in order to compute the statistical significance, the normalization with respect ν is not a necessary step.

Definitions 5-7 are made without any reference to a particular type of system. In their original papers, Edelman and Tononi considered fluctuations around a stationary state of a neural system. Here this measure will be applied to time series[2] of data generated by a dynamical model. In general, these data lack the stationary properties of fluctuations around a fixed point. Moreover, depending upon the case at hand, either transients from arbitrary initial states to a final attractor, or collections of attractor states can be considered (see section 3 and 4). For reasons that will be discussed below, in some cases responses to perturbations of attractor states will also be considered (sections 5 and 6). In all these cases we will use Eq.5, that will therefore be called the *Dynamical cluster index* (DCI).

---

[2] indeed, the method is even more general, see section 7 for a comment

The search for candidate relevant subsets of a dynamical system by means of the DCI requires first the collection of observations of the values of some variables at different instants. It is important to emphasize that no other information is needed; therefore, the method can be applied even if nothing is known about the structure of the system, e.g., (functional) relations among its variables. It is not even necessary to know the values of all the important variables, although of course the quality of the results can be negatively affected by lack of information. In any case, the information provided by this analysis can be complemented with other kinds of knowledge on the system, as discussed in section 7.

In order to find candidate RSs, in principle all the possible subsets of system variables should be considered and their DCI computed. In practice, this procedure is feasible only for small-size subsystems in a reasonable amount of time. Due to this combinatorial factor, some simplifications of the method are required in order to address the study of large-size systems. A simple approximation consists in sampling the configurations, instead of exhaustively analyze them all. In absence of specific heuristics guiding the sampling, uniform sampling can be chosen. More precisely, a maximum number of samples to be evaluated per subsystem size is defined. In this way, for subsets of small size the error introduced by sampling is rather limited. Indeed, the information provided by RSs of small size is usually more significant than that of large-size because large candidate RSs can be composed on the basis of smaller ones.

Nevertheless, a uniform sampling might miss subsets with high DCI value. To overcome this problem, random sampling is complemented with a simple heuristic search. Once all the samples of size k are evaluated, the samples of size *k*+1 are composed of random samples plus all the (*k*+1)-size neighbors of the *k*-size subset with the highest DCI value. This heuristic has proven to be quite effective, because usually the subsets with highest DCI value are composed of subsets which in turn have a high DCI value, compared to the subsets of the same cardinality. Once all the samples for size up to *N*-1 are evaluated, the $T_c$ is computed so as to rank the candidate RSs.

In our experiments, we always relied on the procedure described above. However, other search procedures can be adopted. As an ongoing work, we are experimenting a genetic algorithm for searching the candidate RS with highest DCI value for each size.

In the following we will show the result of the application of this ranking method to some interesting classes of systems. The method draws our attention on the subsets that are highly functionally correlated and that could represent possible candidates relevant subsets.

The data will be presented in a way that is clearly exemplified by the first case studied (see fig. 3.1). In each table, a horizontal line is associated to a subset *S* of the whole system: the variables are ordered from left to right, and a black square is associated to those variables that are present in the subset considered, while a white square to those variables that are not members of that subset. So for example the first row of fig. 3.1d describes a subset composed by the last three variables only (N4, N5 and N6) while the third row refers to a subset composed by N2 and N4, N5 and N6. For obvious reasons, we refer to these rows as "masks": each mask is a graphical representation of a subset. In each row of Fig.

3.2d one also finds a number, that is the $T_c$ of the corresponding subset. The subsets are listed from top to bottom according to their $T_c$s, in decreasing order.

The DCIs (that are at the basis of the computation of the $T_c$s) are computed on dynamical data whose kind may depend upon the specific case that is considered. We will define which type of data are used in the various applications, and will finally comment on their properties in section 7.

## 3. Results on Boolean networks

The case study presented here consists of three synchronous deterministic Boolean networks (BNs), described in Fig. 3.1. BNs are an important framework frequently used to model genetic regulatory networks [Kauffman, 1993, 1995], also applied to relevant biological data [Serra et al. 2004] [Shmulevich et al. 2005] [Villani et al. 2007] and processes [Serra et al. 2010] [Villani et al. 2011]. The aim of this case study is to check whether DCI analysis is capable of recognizing special topological cases, such as causally independent subnetworks. In particular, BN1 is made of two independent components, while in BN2 the same components do interact. Finally, BN3 is the union of the two previous cases.

This system has already been described in [Villani et al. 2013] however the main results are reviewed in this section, as they illustrate the method and provide information about its performance. Note that the behavior of each node is affected by the values of more than one other nodes, so traditional analyses based on correlation between pairs of variables might fail. For example the computation of Pearson correlation coefficients of the networks of this section does not lead to identify related variables, given that only diagonal elements take non negligible values.

The data used in order to determine the relative frequencies that are necessary to compute the entropies required to apply the DCI method are obtained from the states of the various attractors, each one weighted according to its basin of attraction[3]. In order to visualize the procedure, imagine to list all the states of all the attractors in horizontal lines, one below another; each horizontal line contains the values, ordered from left to right. If all the basins of attraction have the same size, each attractor is listed only once, otherwise an attractor is repeated a number of time proportional to its relative frequency. Suppose that this procedure gives rise to a series of $T$ horizontal stripes. Then the relative frequency of, say, value 1 of e.g. the third variable is simply given by the number of "1"s in the third vertical stripe divided by $T$ (of course, the relative frequencies of combinations of values, that are needed to compute the joint entropies, are computed in much the same way.

---

[3] In case of not deterministic or noisy systems a possibility to estimate the attractors' weights is that of using the persistence time of the systems in each of them [Villani and Serra, 2013]

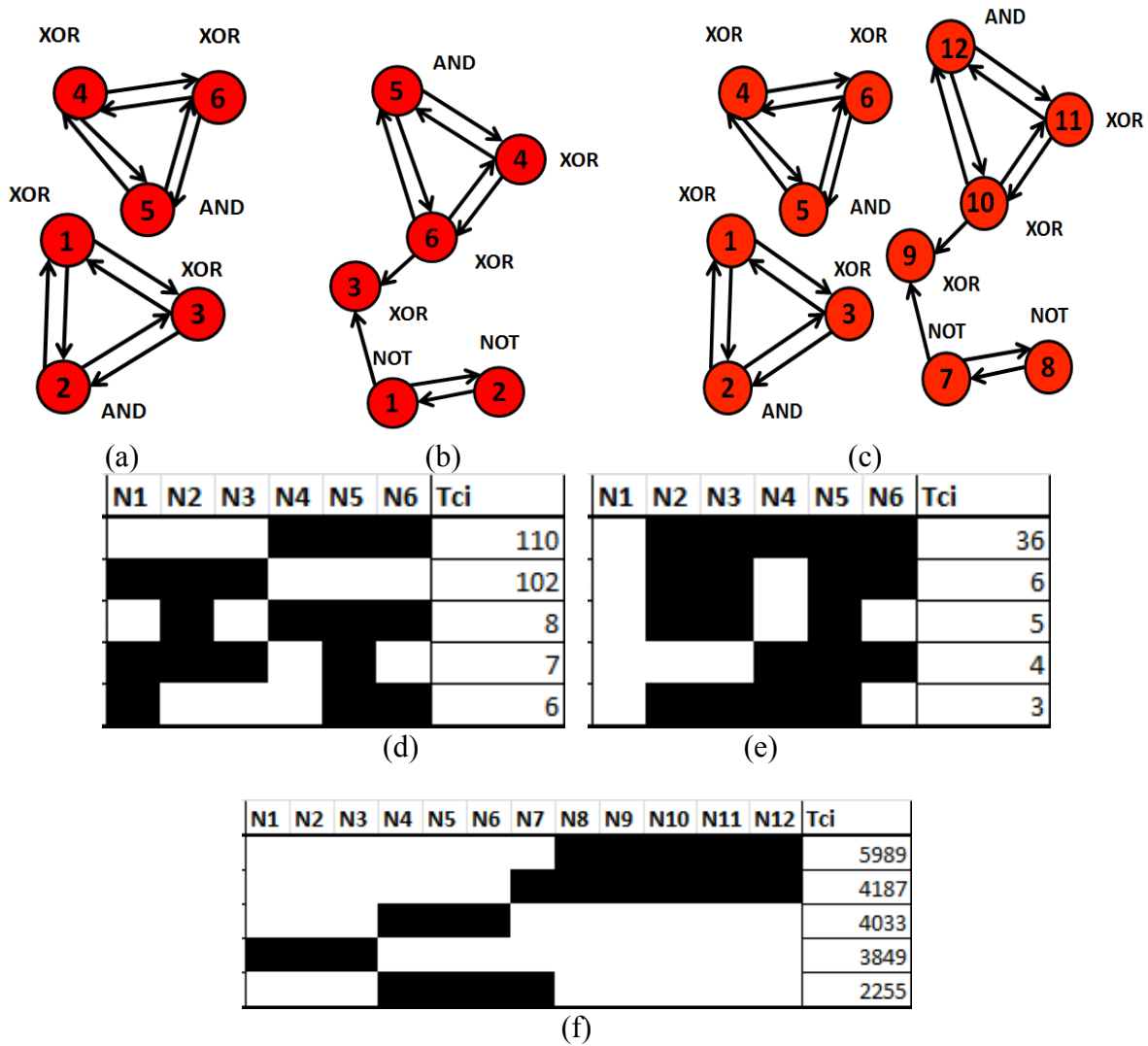

**Figure 3.1** (a) independent Boolean networks (BN1); (b) interdependent networks (BN2); (c) a system composed by the union of both the previous networks (BN3). The boolean function associated to each node is shown close to the node itself. the node is realizing The second part of the figure shows the masks illustrating the elements belonging to the clusters (black on figures) and the corresponding $T_c$ values, for (d) BN1, (e) BN2 and (f) BN3 systems

The rationale for using the attractor states for this analysis is, intuitively, that attractors should be able to capture the important functional relationships in a system. This holds true when the attractor landscape is rich enough, but there are cases where a different approach is needed, as it will be seen in section 5. Further comments on this issue are deferred to section 7.

The DCI analysis is able to correctly identify the two separated subnetworks of BN1 (see the masks of the first and second rows of fig. 3.1d). The same type of analysis clusters together 5 of 6 nodes of BN2: those already clustered in BN1, plus nodes 2 and the node that computes the XOR of the signals coming from the two just mentioned groups. Indeed, all these nodes are needed in order to correctly reconstruct the BN2 series. The analysis is able to identify all MDSs also when all the series are merged together (figure 1f, where the top two clusters correspond respectively to the 5 nodes already recognized in BN2 and to the whole BN2 system, while the third and fourth rows correspond to the independent subgraphs of BN1 - see [Villani et al., 2013] for details. Experiments performed using asynchronous update yielded essentially the same results. Therefore we come to the conclusion that the DCI provides useful information for identifying the relevant subsets in this case (that is admittedly somewhat artificial, but this case was planned in order to test the method itself).

**4. Results on leader-follower dynamics**

To assess the applicability of the method, we studied a simple model in which the integration among variables in a subsystem under observation and its mutual information can be tuned by acting on two parameters. The model abstracts from specific functional relations among elements of the system and could resemble a basic leader-followers model (see Figure 4.1). The system is composed of a vector of n binary variables $x_1, x_2, ..., x_n$, e.g., representing the opinion in favor or against a given proposal. The model generates independent observations of the system state, i.e., each observation is a binary n-vector generated independently of the others, on the basis of the following rules:

- Variables are divided into two groups, $G1 = [x_1,...,x_k]$ and $G2 = [x_{k+1},...,x_n]$
- $x_1$ is called the *leader* and it is assigned a random value in $\{0,1\}$
- the value of the *followers* $x_2,...,x_k$ is set as a copy of $x_1$ with probability $1-p_{noise}$ and randomly with probability $p_{noise}$
- the values of elements of G2 are assigned as a copy of a random element in G1 with probability $p_{copy}$, or a random value with probability $1-p_{copy}$

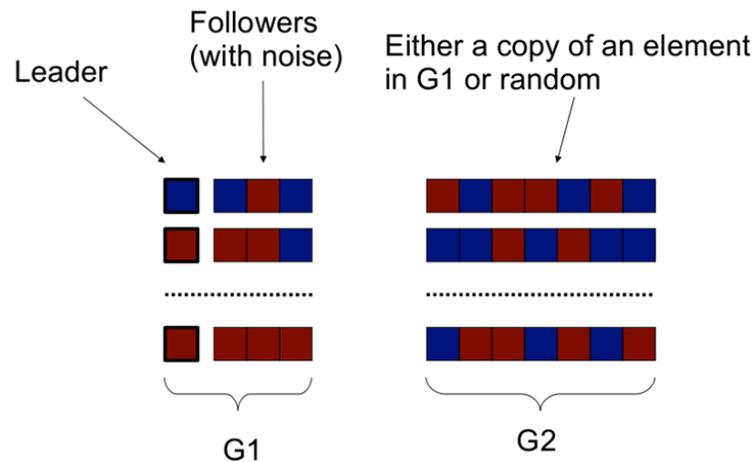

**Fig. 4.1**. Pictorial view of the data generated according to a tunable leader-followers model. Each line corresponds to one observation of the values assumed by system's variables and it is generated independently of the other lines. The leader assumes value 1 with probability 0.5; the followers are assigned the same value of the leader with probability $1-p_{noise}$ and a random value with probability $p_{noise}$. Finally, the other elements copy a random value in G1 with probability $p_{copy}$ and are randomly assigned with probability $1-p_{copy}$.

It is possible to tune the integration among elements in G1 and the mutual information between G1 and G2 by changing $p_{noise}$ and $p_{copy}$. Note that, given significant level of integration, we have two notable cases:
1. MI ≈ 0 → isolated (possibly integrated) relevant set;
2. MI » 0 → integrated and segregated cluster.

The possible scenarios which can be obtained by tuning $p_{noise}$ and $p_{copy}$ can be conveniently illustrated by a 3-dimensional plot. Figure 4,2a shows the behavior of integration of G1 as a function of $p_{noise}$ and

$p_{copy}$. We can observe that it is a decreasing function of $p_{noise}$, while it is independent of $p_{copy}$ (by definition, indeed). The behavior of the mutual information between G1 and G2 is depicted in figure 4.2b; as we can observe, MI increases fast with $p_{copy}$, as this parameter increases the correlation between variables in G2 and G1. Moreover, it also increases with $p_{noise}$, but the reason is that the correlation among variables in G1 increases the randomness of variables in G1, which than behaves similarly to the variables in G2. This last observation deserves to be discussed in more detail, because it concerns one of the main characteristics of the computation of DCI. By definition, the entropy values are computed on the basis of the occurrence of combination of symbols, hence the information conveyed relates to frequencies and not to causal relations. For this reason, integration and mutual information, and consequently DCI, capture relations among the distributions of the values assumed by state variables. This property has the advantage of making it possible to find relevant sets even in the absence of any knowledge on the relations between system's variables. Yet, it has the drawback that it might erroneously detect relations among independent variables drawn from the same distribution or it might miss to identify a relevant set because all the combinations of its variables appear with the same frequency as in a random sequence. However, since dynamical systems of interest are usually dissipative, this circumstance is not so frequent.

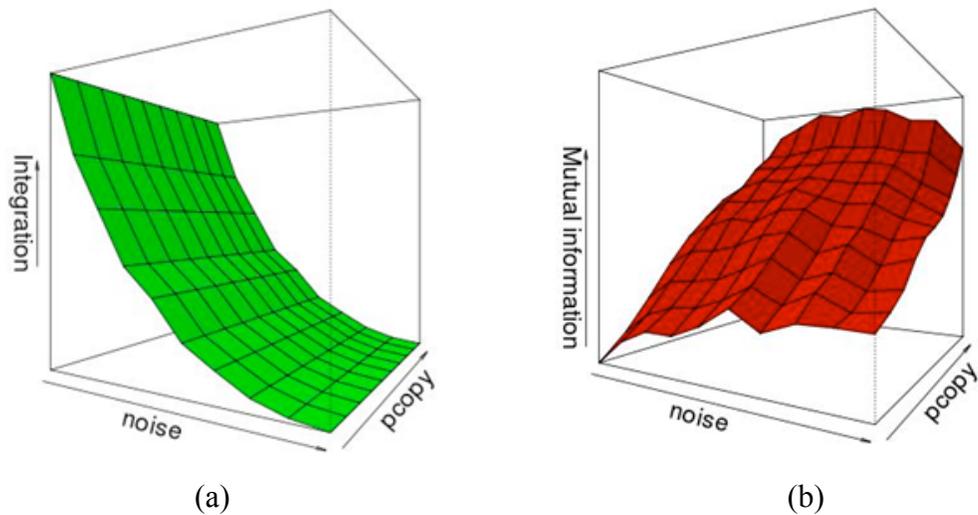

(a)            (b)

**Fig. 4.2**. (a) Integration of G1 and (b) mutual information between G1 and G2 as a function of *pnoise* and *pcopy*. Note that the integration is a decreasing function of $p_{noise}$ while it is independent of $p_{copy}$ (by definition, indeed), whereas mutual information increases more rapidly with $p_{copy}$ than with $p_{noise}$ (see text for explanations)

The special case of $MI \approx 0$ corresponds to the situation in which G1 is almost completely independent of G2 and can be easily detected by observing only *MI*. However, the general case of interest is that of discovering G1 as significant RS, then we would consider cases in which both I and MI are significantly high. In figure 4.3a the matrix representing the cases in which the method correctly detected G1 is depicted. For each value of $p_{noise}$ and $p_{copy}$ in {0.0,0.1,...,1.0} we run the analysis and checked if the RS with highest $T_c$ value corresponded to G1. Dark cells denote the successful cases. As it can be observed, G1 is identified up to $p_{noise}$ equal to 0.8, at decreasing values of $p_{copy}$. The method is quite robust, as it is able to distinguish the group of (even highly) noisy followers from group G2, even at non negligible $p_{copy}$ values.

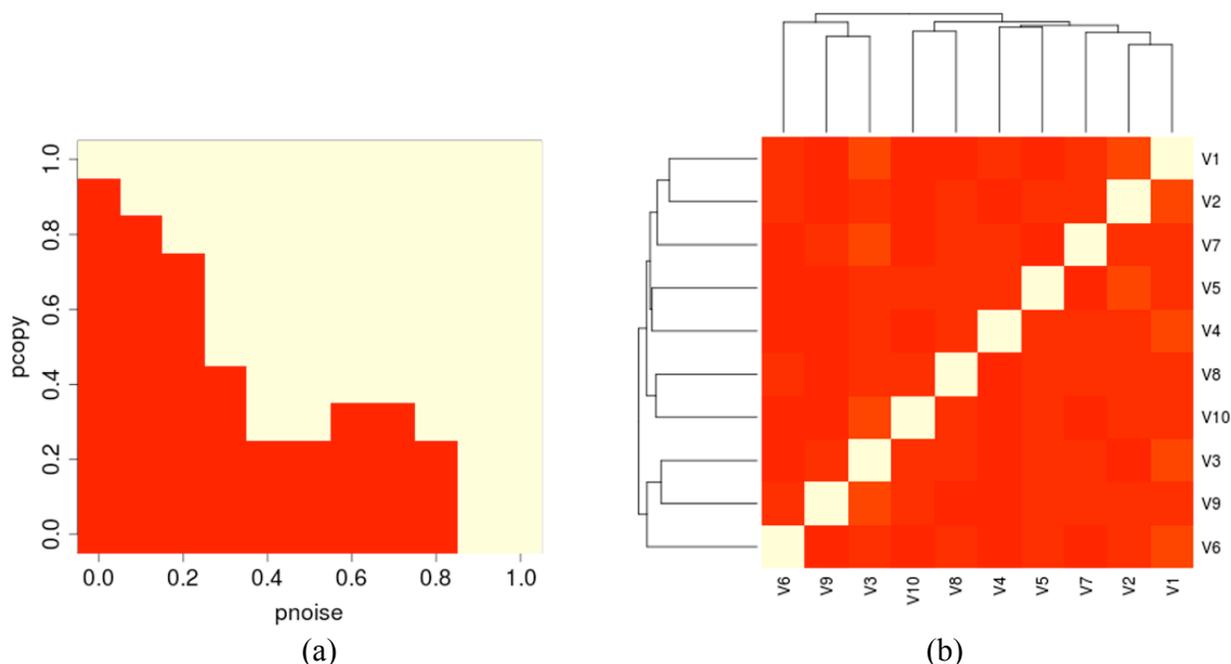

(a)                               (b)

**Fig. 4.3**. (a) Matrix representing the cases in which group G1 is found as the RS with highest $T_c$. Dark red entries denote successful cases. (b) Heatmap with dendrogram representing the result of hierarchical clustering of the variables based on pair correlations. The analysis is made on an instance with $p_{noise}$=0.8 and $p_{copy}$=0.2. The matrix entry $(i,j)$ is the Pearson correlation of variables $V_i$ and $V_j$, computed on the basis of the frequencies of their values; note that G1 is not identified.

Furthermore, it should be observed that the method is superior to hierarchical clustering bases on pair correlations. As an example, in figure 4.3b we show the result of the application of a hierarchical clustering algorithm for the case $p_{noise}$ =0.8, $p_{copy}$=0.2; we can note that G1, i.e., the cluster composed of variables $V_1$,$V_2$ and $V_3$, is missed. The reason of this difference is to be found in the fact that the method based on DCI takes into account multiple and not just pair correlations.

## 5. Results on catalytic reactions

The formation of sets of molecules able to collectively self-replicate is thought to be fundamental both for the origin of life[Eigen and Schuster, 1977a] [Eigen and Schuster, 1977b] [Dyson 1985] [Kauffman, S.A., 1993] [Jain and Krishna, 1998] [Carletti et al., 2008] [Filisetti et al., 2010] [Ruiz-Mirazo et al. 2014] and for new foreseeable biotechnological systems [Solè et al., 2007].

Accordingly, several attempts have been made to identify dynamical cores of self-replicating structures [Farmer et al., 1986] [Kauffman, 1986] [Hordijk et al., 2010], that can be regarded as relevant sets, according to our terminology. All these methods rely on the topological and structural properties of the reaction networks they deal with, hence dynamical information are not taken into account. In this section we investigate the capability of the DCI method to identify CRSs in systems where several reactions take simultaneously place, using only data concerning the values of the system variables (i.e. the concentrations of various species) without any prior knowledge of the reaction graph. As discussed in section 1, we will actually use a dynamical model, well-known to us but totally unknown to the "analyst", in order to generate the data.

In particular, we consider here a relatively simple reaction system with a clear organization (see fig. 5.1). It is composed of two distinct reaction pathways, i.e. a reactions chain (CHAIN from now on) and an autocatalytic set (ACS from now on) that take place in the same vessel, without however directly interfering with each other. The goal of this test will be achieved if the DCI method correctly identifies these subsets. Note that a simplified study of this system has already been presented in [Villani et al., 2013], but the analysis presented here is more complete as it takes into account all the system variables.

The main entities of the model are "polymers" composed of symbols from a binary alphabet {A,B}. The only reactions allowed are condensations, where two polymers are linked to form a longer one, and cleavage, where a polymer is broken in two fragments. Moreover, the only reactions that are assumed to take place at a significant rate are those that are catalyzed. Therefore it turns out that condensations are three-molecular reactions (two substrates and a catalyst): in order for them to happen at an appreciable rate, it is supposed that a temporary complex with a finite lifetime is formed between the catalyst and one of the substrates – the complex can then give rise to the final product if it meets the other substrate. See [Filisetti et al., 2011a] [Filisetti et al., 2011b] for further details.

The two reaction pathways are composed of 6 different condensation reactions, three for CHAIN and three for the ACS reactions (see fig. 5.1). Moreover, a control non-reacting species BB has been added to the system.

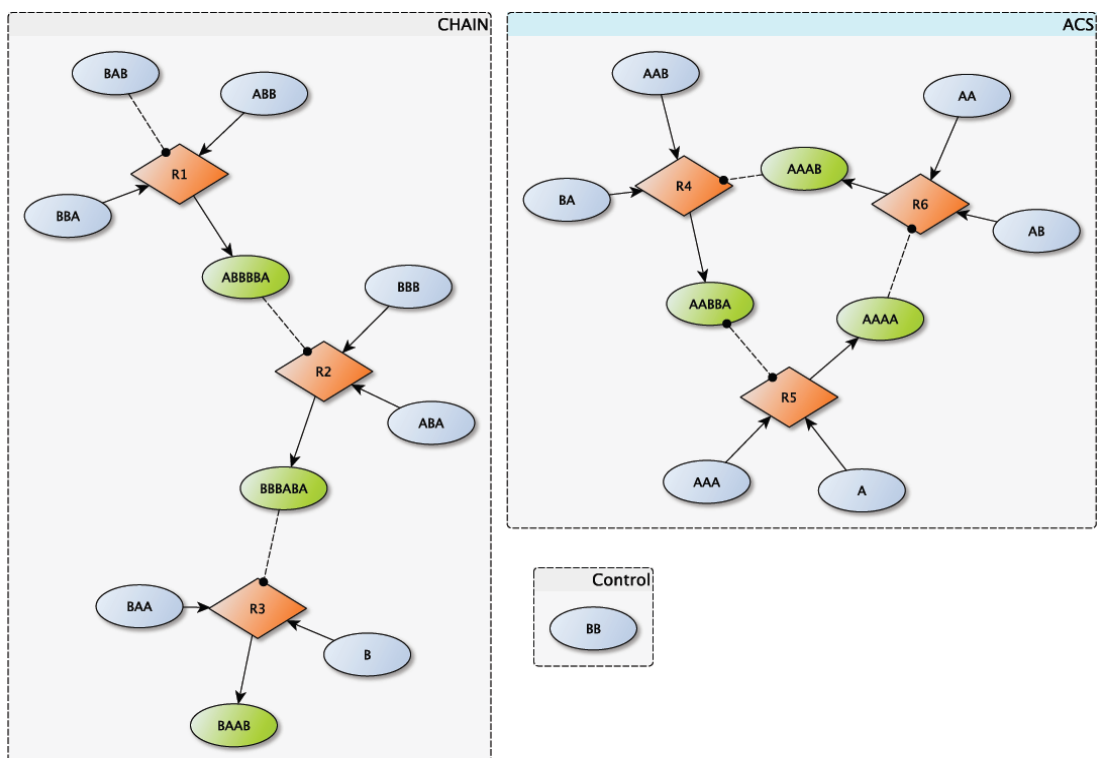

**Figure 5.1** The reaction graph is composed by two co-occurring sets of reactions, CHAIN and ACS. Circular nodes represent chemical species; in particular the blue ones stand for species contained within the incoming flux (food species) while the green ones represent species produced by the activity of the system itself in a CSTR. Diamonds represent reactions, where incoming arrows go from substrates to reactions and outgoing arrows go from reactions to products. Dashed lines indicate the catalytic role of a particular molecular species for a specific reaction.

The model considered here describes the dynamics of the reaction system of Fig.5.1 in a well-stirred tank reactor (CSTR) with a constant influx of food molecules and a constant outflow rate, as the one studied by [Filisetti et al. 2011a] [Filisetti et al. 2011b] [Filisetti et al. 2011c]. While that model was truly stochastic, for the purpose of testing the DCI method we resort here to a simplified deterministic approach, where the reaction scheme is translated, by using the law of mass action, in a set of ordinary differential equations. Simulations are performed using the CellDesigner software [Funahashi et al., 2003] [Funahashi et al., 2008].

It turns out that the asymptotic behavior of the system is a stable fixed point. When this is the case, the DCI method cannot be applied to the system variables, since there would be no sequence of states to consider, but only one. However, in order to display the main features of the system organization, we can perturb the stationary state and study its relaxation behavior. The particular kind of perturbation that has been studied is described in the caption of fig. 5.2, where also the time behaviors are shown.

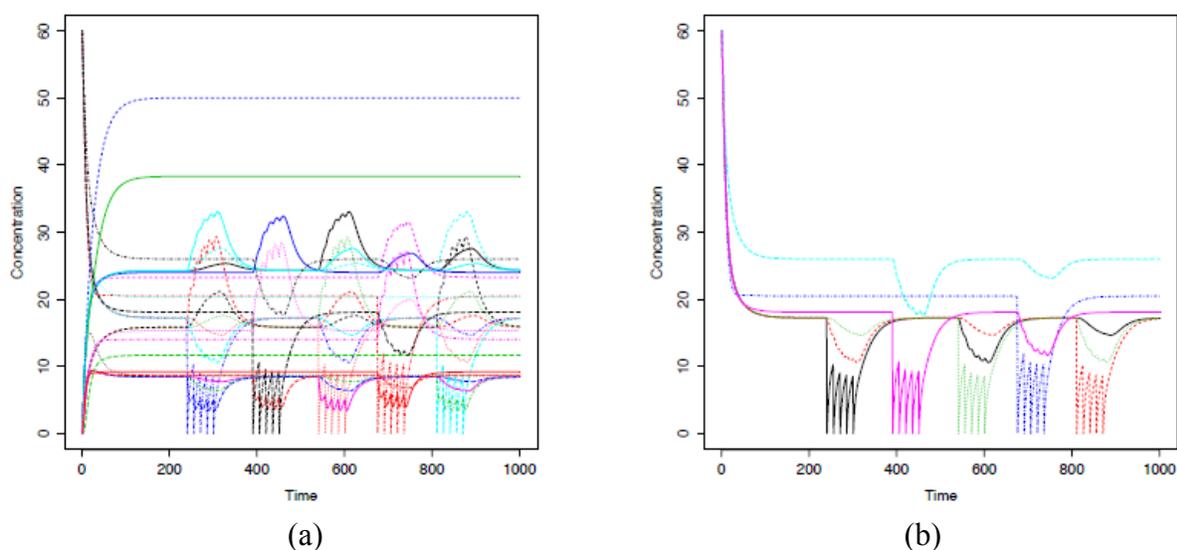

(a) (b)

**Figure 5.2**: The figures show the dynamical behavior of the system in time. Five distinct perturbations have been applied, that affect the molecules ABBBA, BBBABA, AABBA, AAAA and AAAB (whose concentrations are abruptly set to 0 five times each). Then the system is let to relax to his stable fixed point before the next perturbation. In (a) the behavior of all the 26 molecules composing the system; in (b) the behavior of the 6 species that do not belong to the inflow. On the x-axis the time is represented (we take a record every second) whereas concentrations are reported on the y-axis. The kinetic constants of all reactions have the same value (value $k_{dir}$=0.0025 s$^{-1}$mol$^{-1}$); the incoming concentration of each food species is *1.0 mol/s*, whereas the flow rate is such that in each second 2% of the CSTR volume is renewed. The volume of the CSTR is $10^{-18}dm^3$.

Note that the two pathways (CHAIN and ACS) are independent of each other, although they take place in the same environment.

While the dynamical cluster index method might be generalized to continuous-valued variable, in this paper we concentrate on discrete values; therefore continuous concentration values are coded according to a three levels code related to the sign of the time derivatives at time *t* ("decreasing concentration", "no significant change", "increasing concentration"). The dynamical cluster index has been applied both on the reduced system composed only of the molecules that are not found in the inflow, i.e. those of fig. 5.2b, and on the complete set of molecules (fig. 5.2a). Results concerning the former case are shown in figure 5.3.

| | AAAA | AAAB | AABBA | ABBBBA | BAAB | BBBABA | Tc |
|---|---|---|---|---|---|---|---|
| | ■ | ■ | ■ | | | | 85.91 |
| | | | | ■ | ■ | ■ | 64.75 |
| | ■ | ■ | ■ | ■ | | | 50.30 |
| | ■ | ■ | ■ | | ■ | ■ | 50.18 |
| | | | | | ■ | ■ | 39.88 |

**Figure 5.3** the five relevant sets with the higher $T_c$ are shown. As in section 3, sets of variables are represented in rows; a dark box indicates that the species belongs to the set, a white box that it does not belong. The two distinct groups of molecules composing the ACS and the CHAIN are present in the first and second position respectively, ranked according to their $T_c$

Without any information about the topology of the system, the dynamical cluster index is able to identify the two sets ACS and CHAIN as the top-ranking ones. In addition, hybrid sets containing molecules from both sets with a relatively high $T_c$ are also found.

In fig. 5.4 the results concerning the overall system with all the 26 variables are shown, and it can be seen that also in this case the two reaction pathways are clearly identified.

| A | AA | AAA | AAAA | AAAB | AAA_AABBA | AAB | AABBA | AAB_AAAB | AA_AAAA | AB | ABA | ABB | ABBBBA | ABB_BAB | B | BA | BAA | BAAB | AA_BBBAB | BAB | BB | BBA | BBB | BBBABA | BB_ABBBB | Tc |
|---|---|---|---|---|---|---|---|---|---|---|---|---|---|---|---|---|---|---|---|---|---|---|---|---|---|---|
| ■ | ■ | ■ | ■ | ■ | ■ | ■ | ■ | ■ | ■ | ■ | | | | # | | | | | | # | # | # | | | | 542 |
| | | | | | | | | | | | | | ■ | # | ■ | ■ | ■ | ■ | ■ | # | # | # | ■ | ■ | ■ | 496 |
| ■ | ■ | ■ | ■ | ■ | ■ | ■ | ■ | ■ | ■ | ■ | ■ | | ■ | # | ■ | ■ | ■ | ■ | ■ | # | # | # | | | | 390 |
| | | | | | | | | | | | ■ | ■ | ■ | # | ■ | ■ | ■ | ■ | ■ | # | # | # | ■ | ■ | ■ | 365 |
| | | | | | | | | | | ■ | ■ | ■ | ■ | # | ■ | ■ | ■ | ■ | ■ | # | # | # | ■ | ■ | ■ | 365 |

**Figure 5.4**: In the figure the five relevant sets with the higher $T_c$ are shown. The two distinct groups of molecules composing the ACS and the CHAIN are present in the first and second position respectively. Relevant sets are depicted by black stripes indicating the present of the species within the set, "#" char means fixed node not included in the evaluation.

With regard to that case, since the initial dynamical transient is not taken into account, the codification procedure returns molecules with a derivative equal to 0 along all the observed timespan. Since the entropy of a single fixed node is equal to 0, the mutual information of the node with respect to the rest of the system is 0 too. Again, the integration of sets composed only of fixed nodes is 0. Thus, the overall system is analyzed by removing those fixed nodes. In particular, the molecules with fixed concentrations are those involved in the first reaction of the CHAIN (BAB, BBA and ABB) and the control molecule (BB). With regard to the formers, since the perturbations occur in the next steps of the CHAIN, the upstream molecules are not affected. On the opposite, the explanation of the latter is straightforward, once that the control molecule reaches its equilibrium concentration (during the initial transient), its concentration remains fixed in time.

## 6. Results on the Mitogen Activated Protein Kinase (MAPK) cascade

In this section we show, and discuss in some depth, the results of the application of the DCI to models of one of the major cellular signal transduction pathways, known as the Mitogen Activated Protein Kinase (MAPK) cascade. This pathway responds to a wide range of external stimuli, triggering growth, cell division and proliferation, and its biological significance is highlighted by the fact that it is evolutionarily conserved, from yeasts to humans. We will not provide here more details on the biological activity of this pathway, referring to the interested reader to [Sarma & Gosh 2012].

These authors also introduce the three models that will be considered in the following analysis. The basic model is composed of reactions that are quite well-established from an experimental viewpoint, and it has the hierarchical structure shown in Fig. 6.1. The three groups of chemicals that have been identified as the core of this three-layered system are the MAPKKK, MAPKK and MAPK kinases (respectively M3K , M2K  and MK for short) [Widemann et al., 1999]. M3K is activated by means of single phosphorylation whereas M2K and MK are both activated by double phosphorylation [Davis, 1993][Huang CYF and Ferrell, 1996][Bhalla et al., 2002]. Parallel to the phosphorylation by kinases, phosphatases present in the cellular volume can dephosphorylate the phosphorylated kinases (Figure 6.1 shows the schema of the MAPK cascade where each layer of the cascade is dephosphorylated by a specific phosphatase). Note that superimposed on the three-layered structure of substrates-product reactions there is the properly called MAPK signaling cascade, a linear chain of catalysis (dashed lines in fig. 6.1) that transmit the external signal from M3K* to MK**.

When the external signal and the concentrations of the phosphatases are kept constant, a top-down reaction scheme as the one described in fig. 6.1 leads to fixed-point solutions. On the other hand, oscillations have been reported in the MAPK cascade [Shin et al., 2009] (as well as in many other biological circuits) and, in order to account for them, [Sarma & Gosh 2012] make use of two models with feedback, described in fig. 6.2. The two variants (called S1 and S2 in what follows) are characterized by a different disposition of the activating of inhibiting interactions among layers – see figure 6.1 for the details. Note that the insertion of these new catalytic actions alters the "layered" structure of the basic model, that is no longer strictly hierarchical. The presence of feedback from the bottom to the top layer gives rise to what is sometimes called a "tangled hierarchy" [Lane, 2005].

These alternative models are hypothetical but grounded on experimental data as well as on knowledge about chemical interactions; we will not enter here a discussion about the merits and limits of the two models, referring again the interested reader to the original paper, but we will take them "for granted" and we will apply our method to test whether it can discover significant CRSs, without any prior knowledge of the interactions, but on the sole basis of the dynamics of concentrations.

We have simulated the Sarma & Gosh models with the CellDesigner package [Funahashi et al., 2003] [Funahashi et al., 2008], keeping the P1, P2 and P3 phosphatases as constant (as they do) and we obtained the same asymptotic states shown by those the authors.

The time data could be obtained in different ways. However, the basic case leads to a fixed point, so it would be useful to use perturbations as it has been done in the case of the catalytic reaction system. We therefore decided to use the perturbation of asymptotic states for all the three models: in particular, we focused our analysis on the kinases and perturbed[4] each kind of kinase 10 times, by leaving the time to the system to find a new stable situation after the change. The stable situations that are reached can show both oscillating (S1 and S2 systems) or constant concentrations (basic system). Here again, as in section 5, the 3-level coding (concentration decrease, no change, concentration increase) is used.

---

[4]More precisely, we made a perturbation each 1000 simulated seconds (out of  100.000 simulated seconds of a very long run, and avoiding its initial transient phases; samples for the analysis are recorded every 10 simulated seconds). For each chemical specie we choose a random value among the admissible ones, and cyclically imposed it in such a way that each avalanche of changes starts with only one perturbed kinase

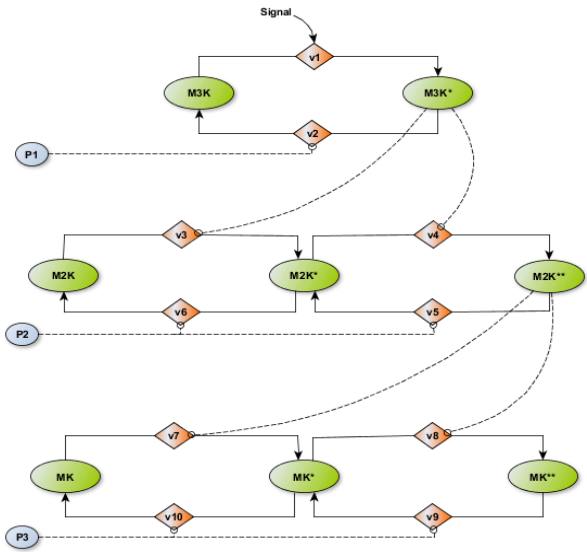

**Figure 6.1: basic model** In the figure the scheme of the three layers MAPK cascade reaction pathway is represented. The "*" symbol stands for the phosphorylation. The signal catalyzes the phosphorylation of M3K to M3K* that in turn catalyzes the phosphorylation of M2K to M2K* and the successive phosphorylation of M2K* to M2K**. Finally M2K** performs the double phosphorylation of MK in MK** that is the final output of the MAPK cascade. P1, P2 and P3 dephosphorylate M3K, M2K and MK kinases respectively. V1-V10 stand for the involved reactions. Dashed lines with circle head represent catalysis; the figure highlights the presence of the three "layers" described on the text

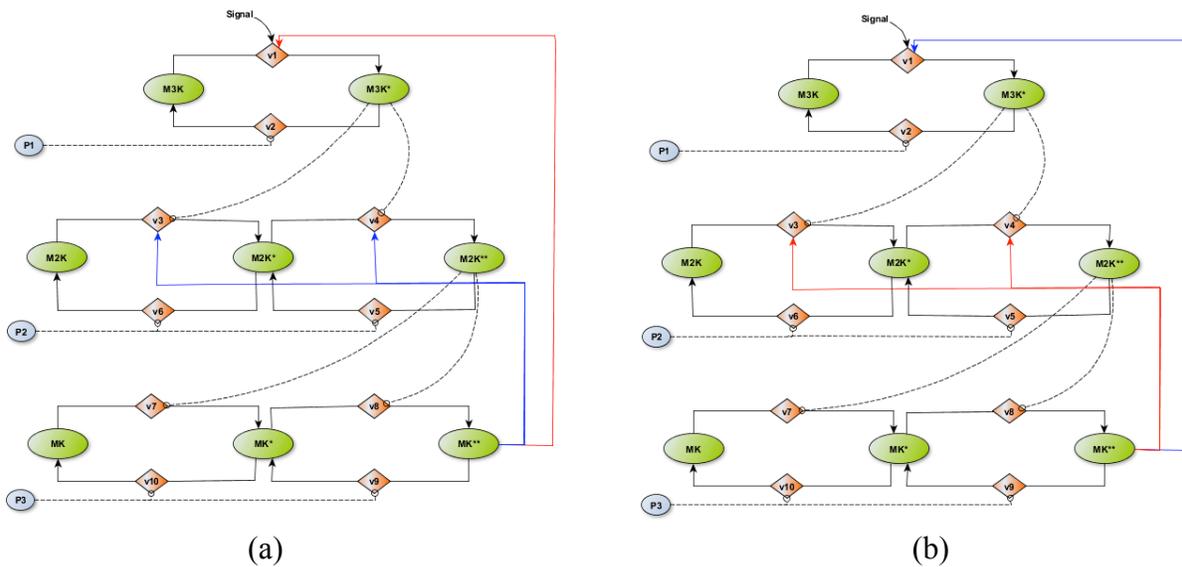

(a) (b)

**Figure 6.2** In the figure the two distinct positive and negative feedbacks added to the basic model are shown. (a) In the left panel (S1 system) the positive feedback goes from MK** to the second layer (M2K, M2K* and M2K**) while the negative feedback goes from MK** to the first layer (M3, M3K*). (b) In the right panel the S2 system with inverted feedback is shown. In S2 the positive feedback goes from MK** to the first layer whereas the negative feedback goes from MK** to the second layer.

As in previous examples the distribution of the $T_c$ values shows that one or a few CRSs outperform the other ones: the higher ranking ones are interesting and are shown in Figure 6.1. The three models of MAPK cascade show both similarities and dissimilarities; so, it is interesting to describe them in detail.

Remarkably our method detects the presence of the "starting" layers (in the following indicated as Layer1, Layer2 and Layer3, ordered from the top to the bottom) among the first highlighted masks; interestingly, at the same time it proposes other groups that well describe the peculiarities of the interactions among the kinases.

## Basic model

| M3K M3K* M2K M2K* M2K** M M* M** | Description | Tci |
|---|---|---|
| | Layer 1 | 2379 |
| | Hole in Layer 3 | 1537 |
| | Layer 2 + Layer 3 | 1429 |
| | Layer 2 | 1189 |
| | Hole in Layer 2 | 1168 |
| | Layer 2 + Intermediate L3 | 1117 |
| | Hole in layer 3 + Catalyst | 1103 |
| | Layer 3 | 1053 |
| | Layer 2 + Hole in Layer 3 | 1012 |

(a)

**Figure 6.2** The figure shows the masks illustrating the elements belonging to the clusters (black on figures) and the corresponding $T_c$ values, for (a) the basic system, (b) S1 system and (c) S2 system. The three layers presented in figure 6.1 are highlighted with different gray levels, and we added a column with a brief description of each mask (see text for a more detailed explanation)

## Model S1

| M3K M3K* M2K M2K* M2K** M M* M** | Description | Tci |
|---|---|---|
| | Layer 1 | 1645 |
| | Hole in Layer 3 | 751 |
| | Hole in Layer 2 | 744 |
| | Hole in Layer 2 + Hole in Layer 3 | 586 |
| | Layer 1 + Intermediate L2 | 551 |
| | Layer 3 + Hole in Layer 2 | 496 |
| | First in Layer 2 + First in Layer 3 | 468 |
| | First in Layer 2 + Hole in Layer 3 | 466 |
| | Hole in Layer 2 + First in Layer 3 | 451 |
| | Hole in Layer 2 + Last in Layer 3 | 435 |
| | First in Layer 2 + Last in Layer 3 | 434 |
| | Last in Layer 2 + Hole in Layer 3 | 433 |
| | Last in Layer 2 + First in Layer 3 | 428 |
| | Last in Layer 2 + Last in Layer 3 | 399 |
| | Layer 3 | 384 |
| | Layer 1 + Intermediate L3 | 384 |
| | Layer 2 | 322 |

(b)

## Model S2

| M3K M3K* M2K M2K* M2K** M M* M** | Description | Tci |
|---|---|---|
| | Layer 2 | 624 |
| | Layer 1 + Layer 3 | 539 |
| | Layer 1 | 178 |
| | First two of Layer 2 | 168 |
| | Layer 1 + Last of Layer 3 | 148 |
| | Layer 1 + Hole in Layer 3 | 139 |
| | First two of Layer 3 | 134 |
| | Layer 3 | 132 |
| | Hole in Layer 2 | 132 |

(c)

Let us first consider the "basic model" whose analysis is described in figure 6.1a. The top ranking CRS is indeed layer one of the original model, while the third one is the set union of layers 2 and 3. This points to a peculiar feature of the concept of relevant subsets, i.e. that good RSs can be (and are often) made out of smaller subsets that are also relevant. Anyway, layer 2 alone ranks 4[th], immediately after the union of layers 3 and 4. Therefore a careful consideration of masks 1, 3 and 4 would by itself draw our attention on all the relevant layers.

We have so far neglected the candidate relevant subset that ranks second in Table 6.1a, where it is called a "hole": it is essentially the third layer without the central molecule. The reason why this subset ranks so high is essentially a dynamical one, shown in figure 6.3a where it can be seen that there is a strong dynamical correlation between the non-phosphorylated and the fully phosphorylated kinases of the third layer, i.e. MK and MK**, in the transient shown (where one grows the other decreases). On the contrary, MK* has a different behavior, and its derivative changes sign when the other two curves intersect each other. Our three-level coding cannot capture this aspect, therefore the behavior of MK* appears less correlated than the other two. Similar remarks apply to the second layer, thus explaining the presence of other high ranking subsets with "holes".

Summarizing, an analysis of table 6.1a would highlight Layer1, Layer2+Layer3 and Layer2, therefore drawing attention to layer three. The fact that "layer three with a hole" ranks higher than layer three itself requires a more detailed understanding of the system dynamics, but the DCI method would anyway address the analyst to understand the dynamical organization of the system.

When we move to models S1 and S2, the layered structure is affected by the feedbacks. Nonetheless, by looking at Table 6.1b, referring to model S2, one can see that the basic level structure is still clearly seen, the three top-rankings being Layer2, Layer1+Layer3, Layer1 alone. The fact that Layer2 is now in a privileged position can be understood as it is now the layer subject to negative feedbacks and it has therefore a strong influence on the system dynamics. We can note the appearance in second position of the association between Layer1 and Layer3 – related to the positive feedback that was added on the basic model. Here we find a new arrangement, called "first two", due to the interactions among the new dynamical structure and a subset of the perturbations: the not phosphorylated form M2K monotonically increases whereas the other two forms show the sequence decrease-increase-decrease (figure 6.3b).

On the contrary, in model S1 the negative feedback acts on layer one, and indeed in this case it is the corresponding mask that has the highest $T_c$, as shown in table 6.1c. The further analysis of the other CRSs is more cumbersome due to the presence of various "holes", therefore we observe that this modification has a stronger influence than S1 on the organization of the reaction network

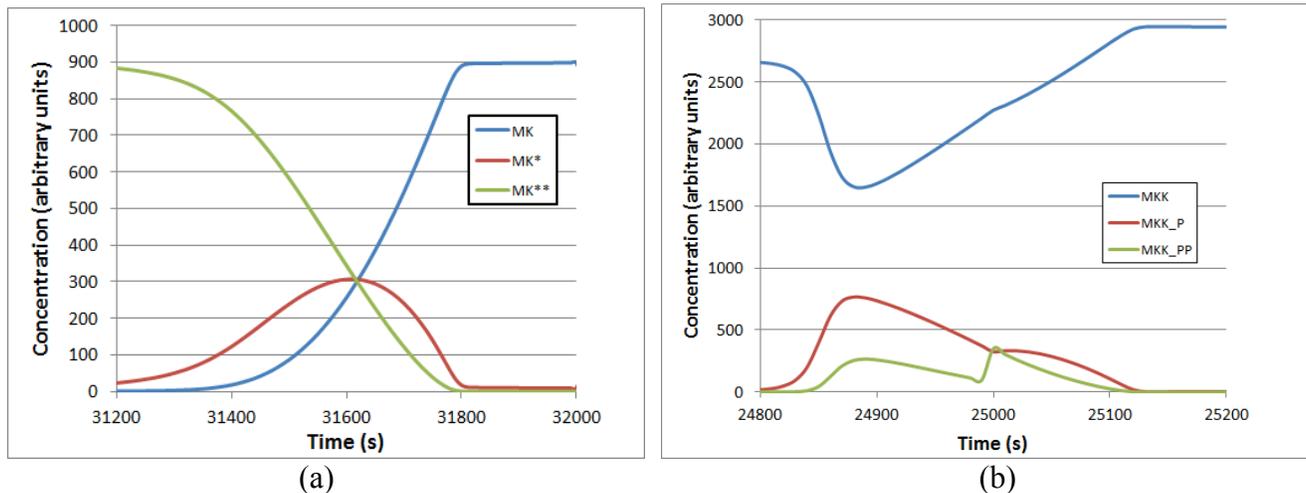

(a) (b)

**Figure 6.3** The concentrations of kinases in Layer3 and Layer2 can show sometimes apparently low correlated behavior (despite these occurrence, these layers are anyhow identified by our method). (a) In basic system, after a perturbation MK monotonically increase and MK** monotonically decrease, whereas MK* shows both behaviors. (b) In S2 system the not phosphorylated form M2K monotonically increases whereas the other two forms show the sequence decrease-increase-decrease.

## 7. Conclusions

In this paper we introduced a method to identify candidate relevant subsets of variables in dynamical systems. The method does not require any knowledge of the relationships among the system variables, but relies on observations of their values in time. We tested its usefulness in a number of different application domains, where the data are generated by a model and the DCI method is applied in order to uncover significant aspects of its organization. In order of increasing difficulty, we considered models of

a) random boolean networks, where the whole network is made of different subnetworks with different topological relationships (independent or interacting subnetworks)
b) leader-follower dynamics, subject to noise and fluctuations
c) catalytic reaction networks in a flow reactor
d) the MAPK signaling pathway in eukaryotes

While we refer to the previous sections for all the details, here we want to synthesize some broad observations and comments that come from those experiences, and point to further research directions.

The main novelty of the present work, in comparison to previous application of the cluster index and of similar measures (Tononi et al. 1998) is that we use it to consider truly dynamical systems, and not only fluctuations around stable asymptotic states. In principle, different kinds of data can be considered. In the case of a deterministic dynamical system, attractors are the main candidates to provide the required time series, and we have shown in case (a) that they can work effectively. Note however that, for reasons discussed in section 5, the method is ineffective in a situation that is sometimes encountered. i.e. if there is just a single fixed point. Therefore we conclude that the use of attractor states only is effective if the attractor landscape is rich enough to show the main features of the system organization. This has usually to be evaluated a posteriori, with the exception of trivial cases like that of a single fixed point.

A possible alternative, or perhaps complementary approach is based on using transient data from arbitrary initial conditions or, perhaps even better, from perturbations of attractor states. This can be very effective, as shown in Sections 5 and 6. Note that the three-level coding used there regards the similarities of the derivatives rather than those of the values of the variables, and that the models that were used in this paper to generate the time series are all based on first-order ordinary differential or difference equations; it should be verified whether the approach is valid also when higher-order dynamical systems are considered. Of course, high-order ODE systems can be transformed in first-order systems by adding variables, but the auxiliary variables that are required might turn out to be unobservable.

When a system is subject to continuous external disturbances the time series directly provides the required data, and the results of section 2 show that our treatment can reveal its organizational features even when a high noise level is present.

Actually, the range of applicability of this method is quite broad, and it needs not necessarily to be limited to dynamical system. Indeed the method needs just a set of frequencies of co-occurences of the values of the system variables. We have always considered here time series data, but this may not always be the case. Note for example that, if we were to scramble the order in which the attractor states of the RBNs of section 3 are reported, this would not change the relative frequencies and therefore the ranking of the relevant subsets. So the method can be applied also to many other systems, since all that is required is a series of "cases" associated to vectors of numerical variables that are not necessarily ordered in time (think for example of different patients, each one described by a vector of values of various symptoms).

A final comment is that the method is not a brute-force one: when there is a clear organization in the system, like e.g. in case (a) when the networks are disjoint, and in cases (b) and (c), the organization can be read out directly from the order list of candidate relevant sets, but this does not always happen. The study of case (d) shows that even in entangled real systems the method provides useful clues to uncover the system organization, but that their interpretation may sometimes require considerable ingenuity.

As far as future directions of research are concerned let us only briefly mention some major ones. Several aspects have been mentioned above, and will be subject to further analysis, including the effective analysis of high-order dynamical systems with a simplified discrete coding and the comparison among different kinds of time series data, i.e. attractors, transients from arbitrary initial conditions and perturbations.

In its present version, the method needs discrete variables, so when dealing with continuous data we have introduced a three-level coding that necessarily misses some information (as documented e.g. by the discussion of section 6). Different coding schemes could be compared, moreover one may also consider the continuous generalization of the proposed method.

Other research will be devoted to improvements of the method: it is apparent that it faces a huge computational problem for large systems, since the number of possible subsets increases extremely fast with the number of variables. Exhaustive search is impossible, and effective heuristics need to be tested and developed to limit the number of candidate subsets to be screened.

Several other specific cases need to be studied to confirm the validity of the approach, mainly from real-world data. Let us mention that, among others, we are considering applications to the study of innovation processes, where data come from the real world, and not from models. In this respect, it is important to realize that the DCI method may be integrated with other approaches: when dealing with real-world problems, and not from data generated by a perfectly known model, a typical situation that is often encountered is that one knows some relationships between variables, but not them all. In this case, the most promising approach would be based on a combination of the a priori knowledge with the DCI, in ways that still need to be tested.

Moreover, it should be recalled that the Dynamical Cluster Index is just one out of several information-theoretic measures that might be applied to analyze dynamical systems. It is worth noticing that the Integration and the Mutual Information can be useful even if used in isolation, and not combined together in the DCI. But other entropies, e.g. those that refer to joint distributions at different times, might prove particularly useful for the study of dynamical systems.

But there is also another line of development of this type of research that may be worth pursuing, concerning the very notion of the organization of a dynamical system. So far, these analyses have largely relied on relatively simple concepts like that of a chain of influences, perhaps with feedback, and that of a layered level structure, perhaps perturbed by some feedback from low levels to higher ones. The case we studied in section 6 shows that there may be some non obvious combinations of variables that seem to play a role, and the DCI could be a way to explore this new territory.

**Acknowledgments**. This article has been partially funded by the UE projects "MD – Emergence by Design", Pr.ref. 284625 and "INSITE - The Innovation Society, Sustainability, and ICT" Pr.ref. 271574, under the 7th FWP - FET programme. Useful discussions with David Lane and Stefano Benedettini are gratefully acknowledged. A reviewed version of this article is currently in press on Artificial Life.

# References

1. Bhalla U.S., Ram P.T., Iyengar R. (2002). MAP Kinase phosphatase as a locus of flexibility in a mitogen-activated protein kinase signaling network. Science, 297(5583):1018–1023
2. Carletti, T., Serra, R., Poli, I., Villani, M. & Filisetti, A. (2008): Sufficient conditions for emergent synchronization in protocell models. *Journal of Theoretical Biology* - Elsevier Ltd. v. 254 pp.741–751
3. Davis J.R. (1993). The Mitogen-activated protein kinase signal transduction pathway. J Biol Chem 268(20):14553–14556
4. Dyson, F. J. (1985). Origins of life. Cambridge: Cambridge University Press
5. Eigen, M., Schuster, P. (1977a). The Hypercycle: a Principle of Natural Self-Organisation, Part A. Naturwissenschaften, 64:541–565
6. Eigen, M. and Schuster, P. (1977b). The hypercycle. A principle of natural self-organization. Part A: Emergence of the hypercycle. Die Naturwissenschaften, 64(11):541–65
7. Farmer, J., Kauffman, S.A., Packard N. (1986) Autocatalytic replication of polymers. Physica D: Nonlinear Phenomena. 220, 50–67
**8.** Filisetti A., Serra R., Carletti T., Villani M., Poli i. (2010) Non-linear protocell models: Synchronization and Chaos Eur. Phys. J. B 77, 249–256
9. Filisetti A., Graudenzi A., Serra R., Villani M., De Lucrezia D., Füchslin R.M., Kauffman S.A., Packard N., Poli I. (2011a) A stochastic model of the emergence of autocatalytic cycles *Journal of Systems Chemistry*, 2:2 doi:10.1186/1759-2208-2-2
10. Filisetti A., Graudenzi A., Serra R., Villani M., Füchslin R.M., Packard n., Kauffman S.A.. Poli I (2011b) A stochastic model of autocatalytic reaction networks *Theory in Biosciences* v.130 Springer Berlin/Heidelberg DOI 10.1007/s12064-011-0136-x
11. Filisetti, A., Graudenzi, A., Serra, R., Villani, M., De Lucrezia, D., and Poli, I. (2011c). The role of energy in a stochastic model of the emergence of autocatalytic sets. In *Advances in Artificial Life ECAL* 2011 Proceedings of the Eleventh European Conference on the Synthesis and Simulation of Living Systems, H. Bersini, P. Bourgine, M. Dorigo, and R. Doursat, eds. (MIT Press, Cambridge, MA), pp. 227-234.
12. Funahashi, A., Tanimura, N., Morohashi, M., and Kitano, H. (2003). CellDesigner: a process diagram editor for gene-regulatory and biochemical networks, BIOSILICO, 1:159-162
13. Funahashi, A., Matsuoka, Y., Jouraku, A., Morohashi, M., Kikuchi, N., Kitano, H. (2008) CellDesigner 3.5: A Versatile Modeling Tool for Biochemical Networks" Proceedings of the IEEE Volume 96, Issue 8, Page(s):1254 – 1265
14. Hordijk, W., Hein J., Steel M. (2010) Autocatalytic Sets and the Origin of Life. *Entropy*. 12, 7, 1733–1742
15. Huang C.Y.F., Ferrell J.E (1996). Ultrasensitivity in the mitogen-activated protein kinase cascade. *Proc.Natl.Acad.Sci.USA* 93(19):10078–10083
16. Jain S., Krishna, S. (1998). Autocatalytic set and the growth of complexity in an evolutionary model. Phys Rev Lett, 81:5684–5687.
17. Kauffman, S.A. (1986) Autocatalytic sets of proteins. J Theor Biol. 119, 1, 1–24
18. Kauffman, S.A., 1993. *The Origins of Order*. Oxford University Press, Oxford.
19. Kauffman, S.A., 1995. *At Home in the Universe*. Oxford University Press, Oxford.
20. Lane D.A. (2005). Level hierarchies in agent-artifact space, in: Pumain D. (ed.), Hierarchies in Natural and Social Systems, Kluver
21. Ruiz-Mirazo K., Briones C., de la Escosura A. (2014). Prebiotic Systems Chemistry: New Perspectives for the Origins of Life Chem Rev. 114(1):285-366
22. Sarma U., Ghosh I (2012) Oscillations in MAPK cascade triggered by two distinct designs of coupled positive and negative feedback loops BMC Research Notes, 5:287
23. Serra, R., Villani, M. & Semeria A. (2004): Genetic network models and statistical properties of gene expression data in knock-out experiments *Journal of Theoretical Biology* 227: 149-157
24. Serra R., Villani M., Barbieri B., Kauffman S.A., Colacci A. (2010) On the dynamics of random boolean networks subject to noise: attractors, ergodic sets and cell types *Journal of Theoretical Biology* - Elsevier Ltd. v.265 pp.185-193
25. Shin S.Y., Rath O., Choo S.M., Fee F., McFerran B., Kolch W., Cho H.K. (2009). Positive- and negative- feedback regulations coordinate the dynamic behavior of the Ras-Raf-MEK-ERK signal transduction pathway. J Cell Science, 122(Pt 3):425–435
26. Shmulevich, I., Kauffman, S.A., Aldana, M., 2005. Eukaryotic cells are dynamically ordered or critical but not chaotic. *Proc. Natl Acad. Sci.* 102, 13439–13444.
27. Solé, R.V., Munteanu A, Rodriguez-Caso C, Macía J. (2007) Synthetic protocell biology: from reproduction to computation. Philos Trans R Soc Lond B Biol Sci. 362, 1486, 1727–1739 (2007)

28. Tononi G., McIntosh A.R., Russell D.P., Edelman G.M. (1998) Functional Clustering: Identifying Strongly Interactive Brain Regions in Neuroimaging Data *NEUROIMAGE* 7, 133–149
29. Villani M., Serra, R., Graudenzi, A. & Kauffman, S.A. (2007): Why a simple model of genetic regulatory networks describes the distribution of avalanches in gene expression data. J. *Theor. Biol.* 249 : 449-460
30. Villani M, Barbieri A, Serra R (2011) A Dynamical Model of Genetic Networks for Cell Differentiation. *PLoS ONE* 6(3): e17703. doi:10.1371/journal.pone.0017703
31. Villani M., Serra R. (2013) On the dynamical properties of a model of cell differentiation *EURASIP Journal on Bioinformatics and Systems Biology*, 2013:4 – Springer
32. Villani M., Filisetti A., Benedettini S., Roli A., Lane D.A., Serra R. (2013) The detection of intermediate-level emergent structures and patterns Advances in Artificial Life, ECAL 13 MIT Press - ISBN: 9780262317092 372-378
33. Widemann C., Gibson S., Jarpe B.M., Lohson L.G. (1999) Mitogen-activated protein kinase: conservation of a three-kinase module from yeast to human. Physiol Rev, 79(1):143–180